\documentclass[%
 reprint,
 amsmath,amssymb,
prl,
]{revtex4-2}

\usepackage{graphicx}
\usepackage{latexsym}
\usepackage[dvipsnames]{xcolor}
\usepackage{epstopdf, epsfig}
\usepackage{caption}
\usepackage{subcaption}
\usepackage{hyperref}
\usepackage{graphicx}
\usepackage{dcolumn}
\usepackage{bm}
\usepackage{amsmath}
\usepackage{amssymb}
\usepackage{placeins}
\usepackage{tikz}
\usepackage{pgfplots}
\usepackage{color}

\begin{document}

\preprint{APS/123-QED}

\title{A unifying mechanism to explain the rate dependent rheological behavior in non-Brownian suspensions: One curve to unify them all}

\author{R. V. More}
\author{A. M. Ardekani}%
 \email{ardekani@purdue.edu}
\affiliation{%
 School of Mechanical Engineering, Purdue University, West Lafayette, IN 47907, USA
}%




\date{\today}

\begin{abstract}
We propose a unifying mechanism based on the Stribeck curve for the coefficient of friction between the particles to capture the shear thinning - Newtonian plateau - shear thickening - shear thinning rheological behavior at low - intermediate - beyond critical - high shear rates, respectively, for a typical dense non-Brownian suspension. We establish the accuracy of the proposed model by comparing the numerical results with experimental data. The presence of non-DLVO (Derjaguin and Landau, Verwey and Overbeek) forces and a coefficient of friction reducing with asperity deformation explain the existence of the Newtonian plateau at the intermediate shear rates and the second shear thinning regime at high shear rates, respectively.

\end{abstract}

\maketitle



\textit{{Introduction.}} Dense suspensions of particles are abundant in nature and industrial applications with examples ranging from household cornstarch solution \cite{dintzis1996shear} to metallic pastes used in solar cells \cite{taylor1981advances}. In spite of the Newtonian behavior of the suspending fluid medium, suspensions exhibit plethora of non-Newtonian behaviors including yield-stress \cite{singh2019yielding}, non-zero normal stress differences \cite{zarraga2001normal}, shear rate dependent rheology \cite{more2020roughness}, and particle migration \cite{pham2015particle} to name a few \cite{stickel2005fluid}. The general consensus amongst researchers is that there is no time scale but a stress scale that gives rise to the non-linear rate dependent behavior in dense particulate suspensions \cite{Guazzelli2019}.

It has been shown that long range interactions like adhesion and repulsion (DLVO forces named after Derjaguin and Landau, Verwey and Overbeek) can lead to shear thinning in suspensions at low shear rates \cite{chatte2018shear,singh2019yielding}. In addition, a theory based on a normal load dependent coefficient of friction has also been put forward to explain this shear thinning behavior \cite{lobry2019shear} in moderately dense non-Brownian suspensions (volume fraction, $\phi < 0.5$). However, these hypotheses cannot explain the shear thickening which is typically observed in dense suspensions. Shear thickening is caused by the transition of the inter-particle contact from lubricated to frictional regimes as the normal force between particle pairs exceeds a threshold value \cite{Mari2014}. This transition triggers force chain networks resulting in a granular flow type behavior in the suspensions at high shear rate values. So, it is clear that the inter-particle interactions such as non-contact repulsive forces and contact forces like friction govern the rheological behavior of non-Brownian suspensions. 

Historically, it has been reported that a typical dense ($\phi \gtrapprox 0.5$) non-Brownian suspension undergoes shear thinning at low shear rates followed by a Newtonian plateau at intermediate shear rates which transitions to shear thickening beyond a critical shear rate and finally the suspension again exhibits shear thinning at high shear rates \cite{hoffman1972discontinuous, stickel2005fluid, chatte2018shear}. Numerical studies to date are able to capture the shear thinning to shear thickening transition at low to intermediate shear rates \cite{Mari2014}. But, there is no general consensus on the mechanism responsible for the shear thinning observed at high shear rates. Over the years, many explanations have been given. These include an increase in the maximum packing density due to breakdown of spanning clusters \cite{nakajima2009rheology}, elastohydrodynamic effects \cite{kalman2008effects}, micro-scale non-Newtonian shear thinning effects of the interstitial solvent \cite{vazquez2016shear}, or inhomogeneous microstructure at high shear rates after the discontinuous shear thickening transition \cite{fall2010shear}. However, none of these explanations are universal. As a result, there is no general theory or model which can capture all the three regimes in the rate dependent rheological behavior typically observed in dense non-Brownian suspensions. In addition, there no explanation for the Newtonian plateau observed before the shear thickening transition in non-Brownian suspensions. In this work, we propose a model based on the Stribeck curve for inter-particle friction along with non-DLVO repulsive forces to quantitatively capture the entire rheological state diagram for non-Brownian suspensions.

The Stribeck curve for friction behavior has been used widely in the literature to explain the sliding phenomenon occurring in lubricated contacts \cite{bayer1994mechanical}. In a typical Stribeck curve, the coefficient of friction, $\mu$, is plotted as a function of the Sommerfeld number, $S= \eta{V}/W$, where $\eta$ is the lubricant dynamic viscosity, $V$ is the relative sliding velocity between contacting surfaces and $W$ is the normal load \cite{bayer1994mechanical}. However, for rough surfaces, the surface asperity height dictates the full-film to boundary lubrication contact transition (see \cite{maru2007consideration} and the references therein). Particle surface roughness is one of the important parameters governing the rheology of dense suspensions as even the most idealized smooth particles have surface irregularities of $O(0.001-0.01)$ times the particle radii \cite{lobry2019shear}. These surface asperities not only lead to inter-particle contacts, but also dictate the friction in interesting ways. Hence, efforts on investigating the influence of particle roughness on dense suspension rheology have gained much traction in the recent years \cite{Tanner2016, more2020roughness, more2020roughness, jamali2019alternative, more2020constitutive}. 



\begin{figure}
\captionsetup{labelfont=bf,textfont=small, justification=raggedright}
\includegraphics[width=0.5\textwidth]{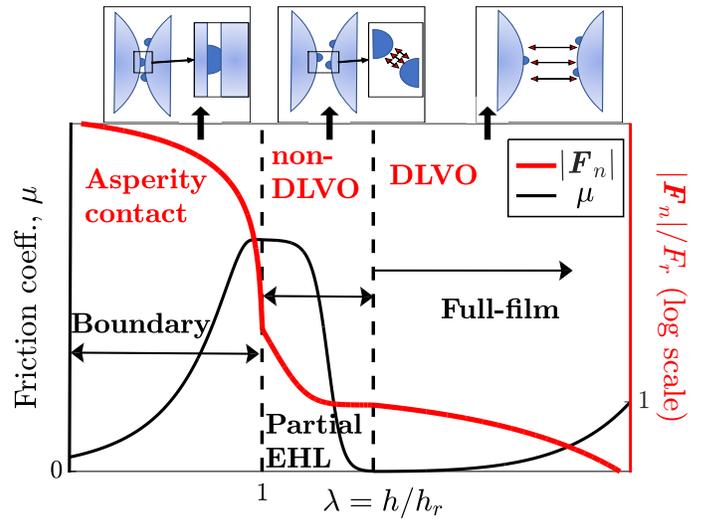}
\caption{\label{fig:fig1}Schematic showing the coefficient of friction, $\mu$ (thin black line), and the dimensionless normal force magnitude, $|\bm{F}_n|$ (thick red line), between a close particle pair as a function of dimensionless inter-particle gap, $\lambda=h/h_r$. Boundary, partial elastohydrodynamic (EHL) and full film lubrication regimes in the Stribeck curve are demarcated based on the value of $\lambda$. Similarly, dominant inter-particle interactions in each of these regimes are also shown in red font. The insets at the top show the various regimes in terms of separation between two close particles. The arrows in these insets are shown to qualitatively indicate the size of the interparticle gap and the range of the dominant inter-particle interaction with respect to the roughness and the particle size.}
\vspace{-5mm}
\end{figure}

In the case of rough particles coming into contact, the average roughness height results in an additional secondary length scale (along with the primary length scale which is particle size) in the system. While the particle size distribution governs the hydrodynamic interactions, the secondary length scale introduces geometrical and inter-particle force constraints \cite{more2020roughness}. So, we define $\lambda$ as the dimensionless gap between the particles, i.e., $\lambda=h/h_r$. Here, $h$ is the inter-particle gap, and $h_r$ is the average roughness height (defined below). High $\lambda$ signifies no dry contact between the particles and the friction force is mostly due to the force transmitted via lubrication interactions (full-film contact) \cite{bayer1994mechanical}. As particles come closer to each other, $\lambda$ decreases, and partial-elastohydrodynamic lubrication (partial-EHL) results in a sudden rise in $\mu$ \cite{hutchings2017tribology, neale1995tribology}. In this regime, partial dry contact between the particles is expected to occur. In addition, as the inter-particle gap becomes comparable to mean particle surface roughness size, repulsive forces of non-DLVO origin (arising when the inter-particle gap is comparable to surface roughness due to hydration or stagnant charge layer on the particle surface) are expected to be present with magnitudes a few orders higher than the repulsive forces of DLVO origin, viz, arising from the double layer potential \cite{adler2001origins, kamiya2008characteristics, diao2016molecular}. As $\lambda$ decreases further, the contact enters boundary lubrication. In this regime, the coefficient of friction has a high value if the contact between the particles is elastic which is true if the asperity deformation is smaller than a threshold value $\delta_c$ \cite{brizmer2007elastic}. If $\lambda$ decreases even further, the contact enters a plastic regime which results in a significant reduction in the coefficient of friction. These phenomena are depicted in fig.~\ref{fig:fig1}. We briefly elaborate on the methods and simulation conditions used in this study in the following section before presenting the main results.

\textit{{Methodology and governing interactions.}} We simulate the shear flow of neutrally buoyant inertia-less bi-spherical particles with radius ratio $1.4$ and equal volume fractions in a cubical domain of size $L=15a$. Here $a$ is the radius of the smaller particle. For this particular particle size distribution, the dry close packing fraction ($\phi_{RCP}$) is 0.66 \cite{more2020constitutive}. We use $\phi_{RCP}$ to normalize the volume fraction ($\phi$) values in this study. Simulation results do not change much for a bigger domain size $L=20a$. The suspending fluid is Newtonian with viscosity, $\eta_0$. The suspension flows under an imposed shear rate $\dot{\gamma}$ with Lees-Edwards periodic boundary conditions at all the sides. Also, the P\'eclet number, $Pe > O(10^3)$ \cite{hoffman1972discontinuous,stickel2005fluid,chatte2018shear}, so, the flow is in the non-Brownian regime. 

We use Ball-Melrose approximation \cite{Ball1997} to calculate the hydrodynamic interactions, $\bm{F}^H$, repulsive force of electrostatic origin, $\bm{F}^r$, Van der Waals attractive force, $\bm{F}^A$, repulsive forces of non-DLVO origin, $\bm{F}^{R}$, and contact interactions, $\bm{F}^C$. The repulsive forces ($\bm{F}^r$ and $\bm{F}^R$) act normally towards the particle center and decay with inter-particle surface separation $h$ over a Debye length $\kappa^{-1}$ as $|\bm{F}^r| = F_r\textrm{exp}(-\kappa{\left( h-2h_r \right)})$ for $h>2h_r$ and $|\bm{F}^r| = F_r$ for $h \le 2h_r$. The non-DLVO repulsive forces are dominant when the inter-particle gap is comparable to particle surface roughness size \cite{parsons2014surface, eom2017roughness}. So, we use a non-DLVO repulsive force for $ h_r \le h \le 2h_r$ with an exponentially decaying form $|\bm{F}^{R}| = F_R\textrm{exp}(-A(h-h_r))$ for $h \ge h_r$ \cite{szilagyi2014polyelectrolyte} and $|\bm{F}^{R}| = F_R$ for $h < h_r$. We choose $A=1000$ for this study. Similarly, the attractive force of Van der Waals origin also acts normally but in the opposite direction to the repulsive force and is modelled as $|\bm{F}^A| = F_A/\left( \left( h-h_r \right)^2 + 0.01  \right)$. We use the DLVO repulsive force as the characteristic force scale to non-dimensionalize the governing forces. So, the characteristic stress scale is given by $\sigma_0 = F_r/6\pi{a}^2$ (and rate scale, $\dot{\gamma}_0=\sigma_0/\eta_0$), related to the transition from lubricated contacts (hydrodynamic) where particles are separated to touching contacts wherein the surface asperities come into direct contact.

We model the surface roughness as a hemispherical bump of size, $h_r$, on the surface of the base sphere. The contact interaction is modeled using the Hertz law for the normal contact force ($|\bm{F}^{C}_n| = k_n (\delta/\delta_c)^{3/2}$) and a linear spring for the tangential contact force ($\bm{F}^{C}_t = k_t \boldsymbol{\xi}_t$), respectively as done in the discrete element method (DEM). Here, $\delta=h_r-h$ is the asperity deformation, $\delta_c$ is the critical asperity deformation for elastic to plastic transition and $\boldsymbol{\xi}_t$ is the tangential spring stretch. The contact activates only when $h \le h_r$. Contact interactions obey the Coulomb’s friction law, $|\bm{F}^C_t| \leq \mu{\bm{F}^{C}_n}$. The details and validation of the algorithm can be found elsewhere \cite{more2019, more2020roughness, more2020constitutive}. The magnitude of the resultant normal force between a particle pair as a function of inter-particle gap is shown in fig.~\ref{fig:fig1}. It is well known that the coefficient of friction is not constant and depends on the normal load $|\bm{F}^{C}_n|$ between the particles \cite{comtet2017pairwise, chatte2018shear, brizmer2007elastic}. Since $|\bm{F}^{C}_n| \propto \delta$ following the Hertz law, $\mu$ can also be described as a function of the dimensionless inter-particle gap, $\lambda = h/h_r$, (since $\delta = 1-\lambda{h}_r$) between the particles.

\textit{{Friction coefficient.}}  We use the dimensionless gap size ($\lambda = h/h_r$) dependent Stribeck curve to model the coefficient of friction \cite{hutchings2017tribology, neale1995tribology}. For $\lambda > 1$, the reduction in $\mu$ with decreasing $\lambda$ is captured in lubrication interactions \cite{Fernandez2013}. We approximate $\mu$ in the partial-EHL regime by a step function \cite{Fernandez2013} for simplicity. For $\lambda \leq 1$, asperities come into contact resulting in a sudden rise in $\mu$. $\mu$ has a high value if the contact is elastic, i.e., $\delta \leq \delta_c$, where $\delta$ is the asperity deformation defined as $\delta=|h-h_r|$ \cite{brizmer2007elastic, maegawa2015effect}. If the asperities deform further such that, $\delta > \delta_c$, the contacts transition into plastic regime resulting in a steep decrease in $\mu$. Experimental measurements \cite{chatte2018shear} have shown that the coefficient of friction decreases with the normal load as $\mu = -a'*\textrm{ln}(|\bm{F}^C_n|)+b'$ or in terms of $\delta$ we can say, $\mu = -a*\textrm{ln}(\left( |\delta|/\delta_c\right)^{3/2})+b$ where $a'$, $b'$, $a$ and $b$ are constants. We choose $a=1/3$ and $b=-0.2$ in this study. We call this friction model \textit{log decay friction}. In addition, we also test two other forms for the coefficient of friction in the boundary lubrication regime as it might vary depending on the suspension particle material. One of them is the experimentally validated Brizmer model (\textit{Brizmer friction}) to calculate the asperity deformation dependent $\mu(\delta) = 0.27*\textrm{coth}\left(  0.27*((\delta/\delta_c)^{3/2})^{0.35} \right)$ \cite{brizmer2007elastic, lobry2019shear, more2019}. We also test a hypothetical expression for the coefficient of friction given by $\mu = 1-\textrm{exp}(-\lambda^2/(1-\lambda))$ (\textit{exponential decay friction}). These friction models have the same characteristic behavior with $|\bm{F}^C_n|$ or $\delta$ as the \textit{log decay friction}. To keep the numerical costs tractable, we set $\delta_c = 0.05*h_r$ \cite{more2019}. All the important simulation parameters are summarized in table~\ref{tab:table1}. 


\begin{table}
\caption{\label{tab:table1}Simulation parameters}
\begin{ruledtabular}
\begin{tabular}{lccccccr}
$\phi$ &  $\dot{\gamma}/\dot{\gamma_0}$ & ${\kappa}^{-1}$ & $F_A$ & $h_r$ & $\delta_c$ & $F_R$ \\
\hline
$0.52$ \& $0.57$ & $0.001 - 50.0 $ &  $0.04a$ &  $0.001F_r$ & $0.01a$ & $0.05h_r$ & $10F_r$ \\
\end{tabular}
\end{ruledtabular}
\vspace{-5mm}
\end{table}


\begin{figure}[tbh!]
\captionsetup{labelfont=bf,textfont=small, justification=raggedright}
    \centering
    \begin{subfigure}[t]{0.25\textwidth}
        \centering
        \includegraphics[height=0.75\textwidth]{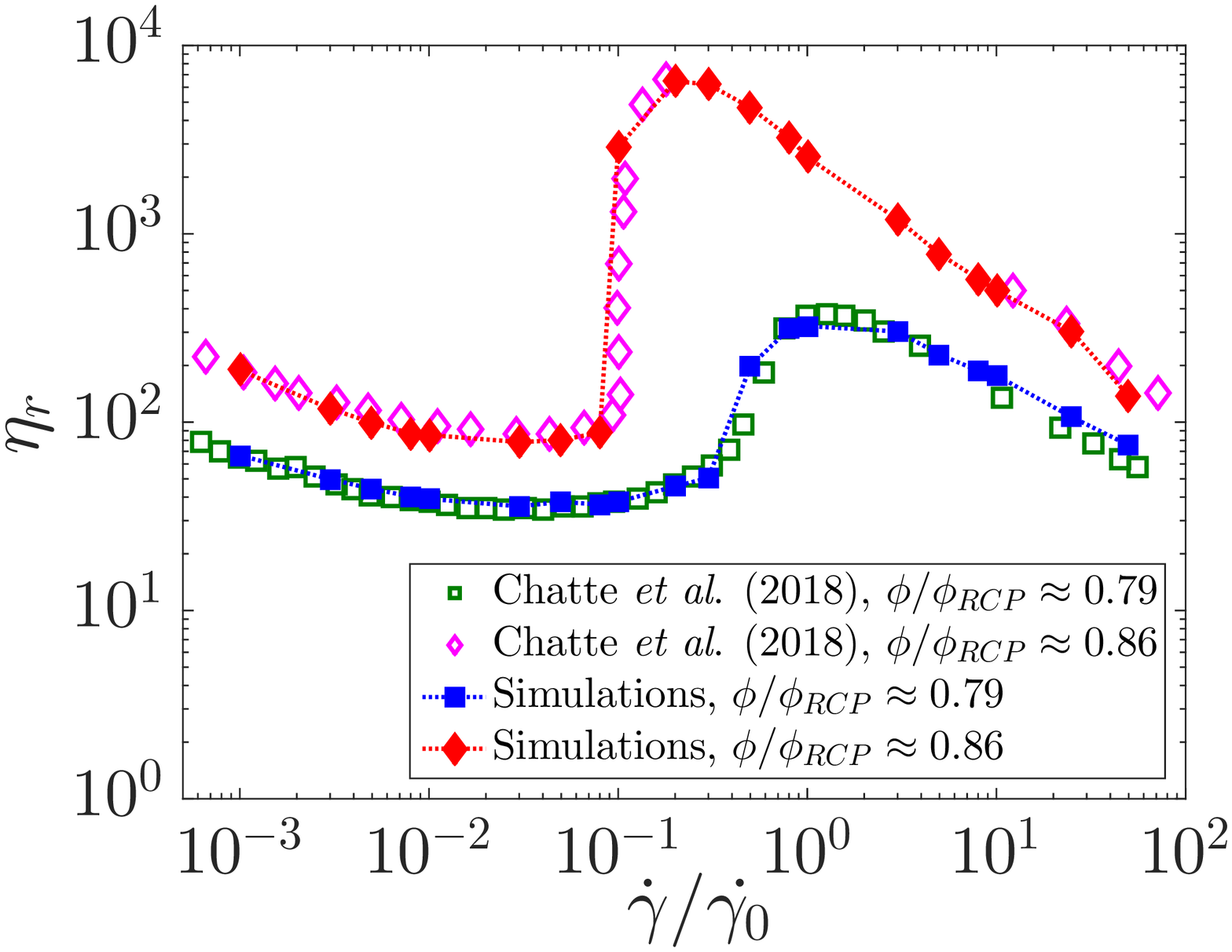}
        \caption{\label{fig:fig2a}}
    \end{subfigure}%
    ~ 
    \begin{subfigure}[t]{0.25\textwidth}
        \centering
        \includegraphics[height=0.75\textwidth]{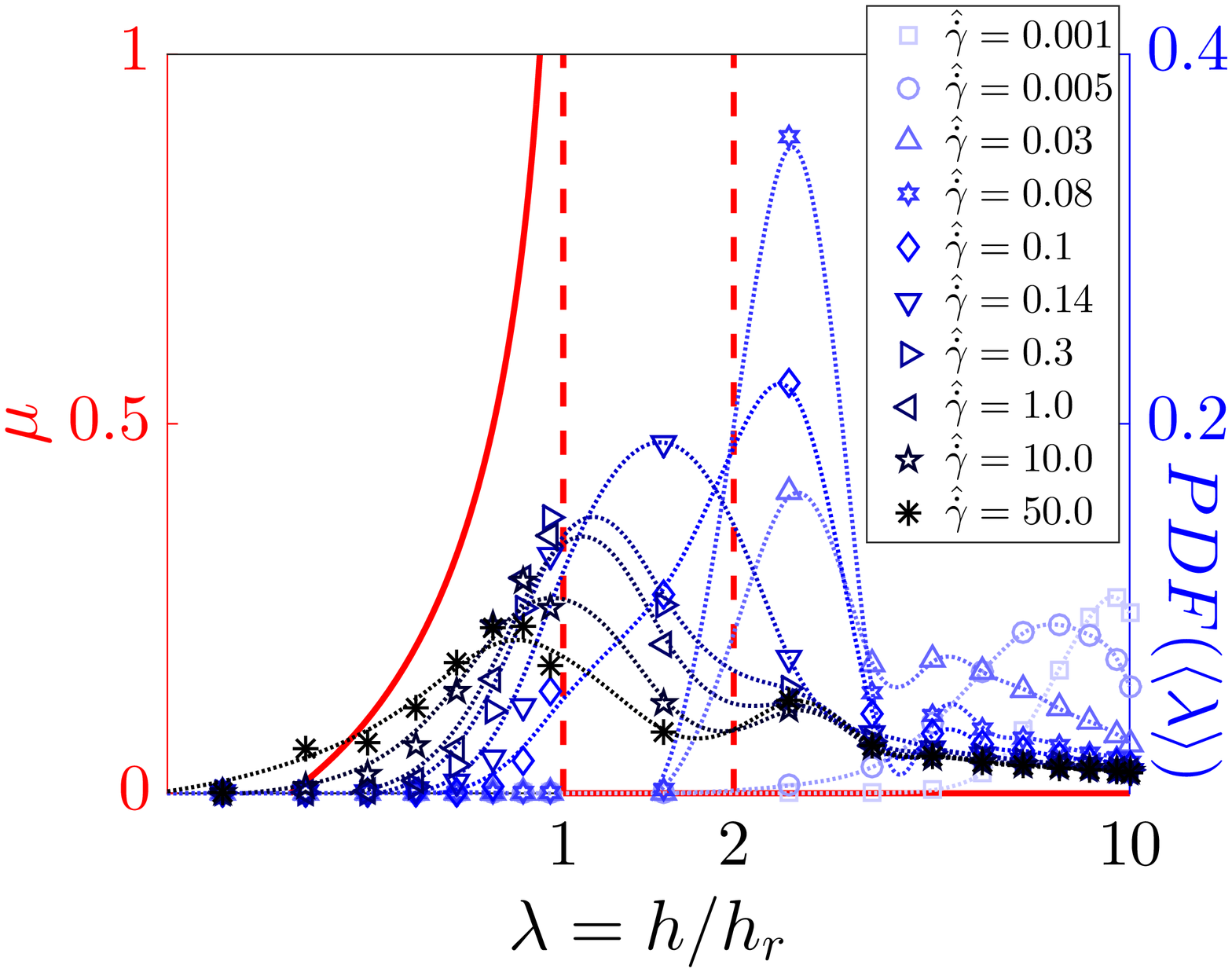}
        \caption{\label{fig:fig2b}}
    \end{subfigure}
    \caption{\label{fig:fig2}Results: a) Relative viscosity as a function of dimensionless shear rate ($\hat{\dot{\gamma}}=\dot{\gamma}/\dot{\gamma_0}$) for two different volume fractions compared against experiments (we choose $\dot{\gamma_0}=200 s^{-1}$ for experimental data) of Chatte \textit{et al}., (2018) \cite{chatte2018shear}. The volume fractions are scaled with dry close packing fraction $\phi_{RCP}$. $\phi_{RCP}=0.66$ for the simulations. b) Probability distribution function (PDF) (right y axis, symbols and dotted lines) of ensemble average of the dimensionless interparticle gap ($\lambda$) with increasing dimensionless shear rates ($\hat{\dot{\gamma}}$) along with the coefficient of friction (left y axis, solid lines) for \textit{log decay friction} model. The peak and mean of the PDF shift to the left on the curve which explains the different regimes observed in (a). Dotted lines are spline fits to the data for guiding the eye. Dashed lines demarcate the transition between interaction ranges as explained in fig.~\ref{fig:fig1}.}
\end{figure}

\begin{figure}
\captionsetup{labelfont=bf,textfont=small, justification=raggedright, singlelinecheck=off}
\includegraphics[width=0.5\textwidth]{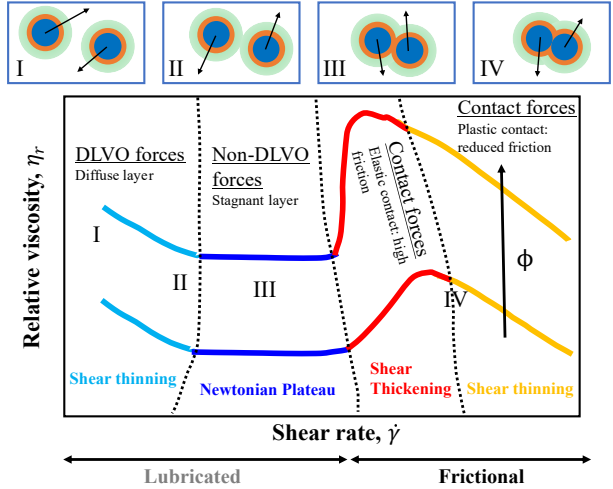}
\caption{\label{fig:fig3}Scheme of the physics involved in the shear thinning ($I - II$) - Newtonian plateau ($III$) - shear thickening - shear thinning ($IV$) regimes in the rheological behavior of a typical dense non-Brownian suspension. The insets at the top show the approximate inter-particle gaps in regimes $I-IV$. In these insets, the outermost circle represents the range of DLVO forces, the inner orange circle represents the range in which non-DLVO forces are dominant and the innermost circle represents the particles. Thus, with increasing shear rate we observe different regimes depending on which forces are dominant in the suspension on average as depicted by the overlaps of different force zones in the insets.}
\vspace{-5mm}
\end{figure}

\textit{{Results \& discussion.}}  We demonstrate the accuracy of the proposed model by direct comparison of the calculated suspension relative viscosity with experimental values for polyvinyl chloride particles suspended in a Newtonian fluid medium \cite{chatte2018shear} in fig.~\ref{fig:fig2a}. The proposed model does an excellent job in quantitatively capturing the rate dependent rheological properties in low, intermediate and high shear rate limits, respectively. This shows that the hypothesis that a universal friction diagram based on the ``Stribeck curve" accurately captures the microscopic competition between different force scales in the system and recovers the transition from shear thinning to shear thickening and then to shear thinning that is typical to dense non-Brownian suspensions is indeed true.

We plot the probability distribution (PDF) of the ensemble average of the dimensionless inter-particle gap $\langle \lambda \rangle$ at different shear rate values corresponding to different regimes in the rheological state diagram (fig.~\ref{fig:fig2b}) to explain the observed shear rate dependent rheological behavior. With increasing shear rate values, the peak and mean of the PDF of $\langle \lambda \rangle$ shift to the left on the Stribeck curve. This determines the various transitions in the rheological state diagram. At low shear rates, the particles are prevented from coming into direct contacts due to the combined effect of the repulsive and attractive forces of the DLVO origin. This is analogous to having particles with bigger radii. As we increase the shear rate, the particles are pushed closer resulting in the reduction of this apparent bigger radius. As a result, the effective volume fraction of the suspension decreases with increasing shear rate in this regime which results in the observed shear thinning. In the intermediate shear rate regime, the stress is high enough to overcome the DLVO repulsive barrier between the particles so that the particles are on average separated by a distance $\approx O(h_r)$. But the stress is not high enough to overcome the short range non-DLVO repulsion which is an order of magnitude higher than the DLVO barrier. This leads to the Newtonian plateau in the relative viscosity. This effect is similar to the effect of Brownian forces in colloidal suspensions in the intermediate  P\'eclet number regime which gives rise to the Newtonian plateau in colloidal suspensions. This plateau in the $\eta_r$ at intermediate $\dot{\gamma}$ values is not present if we do not consider short range repulsive forces of non-DLVO origins \cite{singh2019yielding}. This indicates the governing role of short range repulsive forces in dense non-Brownian suspensions.

If we increase the shear rate further, the stress in the suspension becomes high enough so that the repulsive barrier due to the DLVO and non-DLVO forces breaks and the particles come into contacts due to the touching of asperities on their surfaces. The contact remains in the elastic region resulting in a high $\mu$ between the particles. This leads to a jump in the suspension viscosity. The shear thickening transition takes place above a critical shear rate value ($\dot{\gamma_c}$, e.g., $\dot{\gamma_c}/\dot{\gamma_0}$ for $\phi/\phi_{RCP} \approx 0.86$ is 0.1). In the shear thickening transition regime, the viscosity increases gradually (\textit{continuous shear thickening}) at lower volume fractions while it undergoes a sudden increase (\textit{discontinuous shear thickening}) at higher volume fractions. As we increase the shear rate further, the asperities are plastically deformed  ($\delta > \delta_c$). As a result the coefficient of friction between the particles decreases significantly which gives rise to the second shear thinning transition at high shear rates. The consequences of this shift in the PDF of $\langle \lambda \rangle$ to the left with increasing $\dot{\gamma}$ are depicted pictorially in fig.~\ref{fig:fig3}.

\begin{figure}
\captionsetup{labelfont=bf,textfont=small, justification=raggedright}
    \centering
    \begin{subfigure}[t]{0.25\textwidth}
        \centering
        \includegraphics[height=0.75\textwidth]{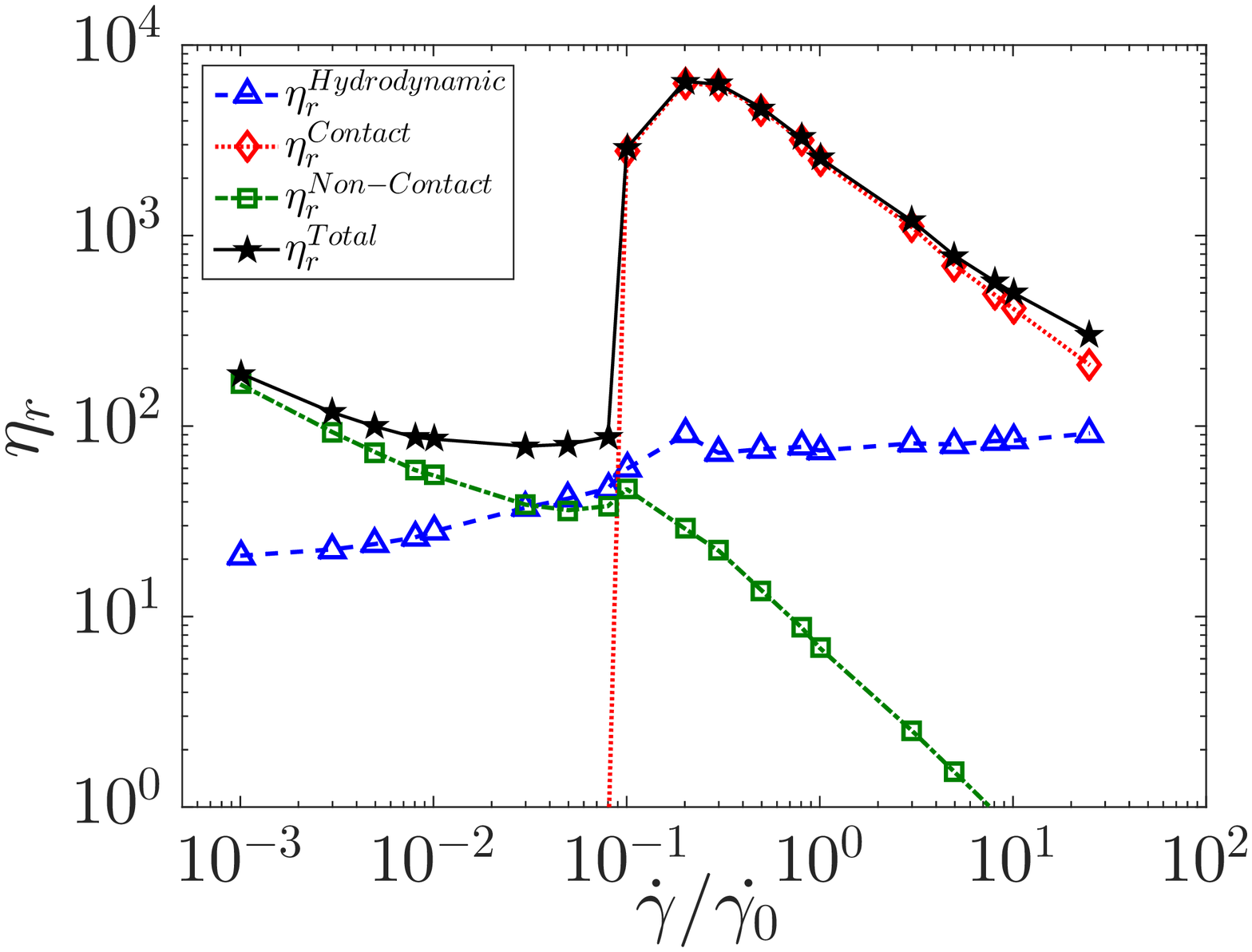}
        \caption{\label{fig:fig4a}}
    \end{subfigure}%
    ~ 
    \begin{subfigure}[t]{0.25\textwidth}
        \centering
        \includegraphics[height=0.75\textwidth]{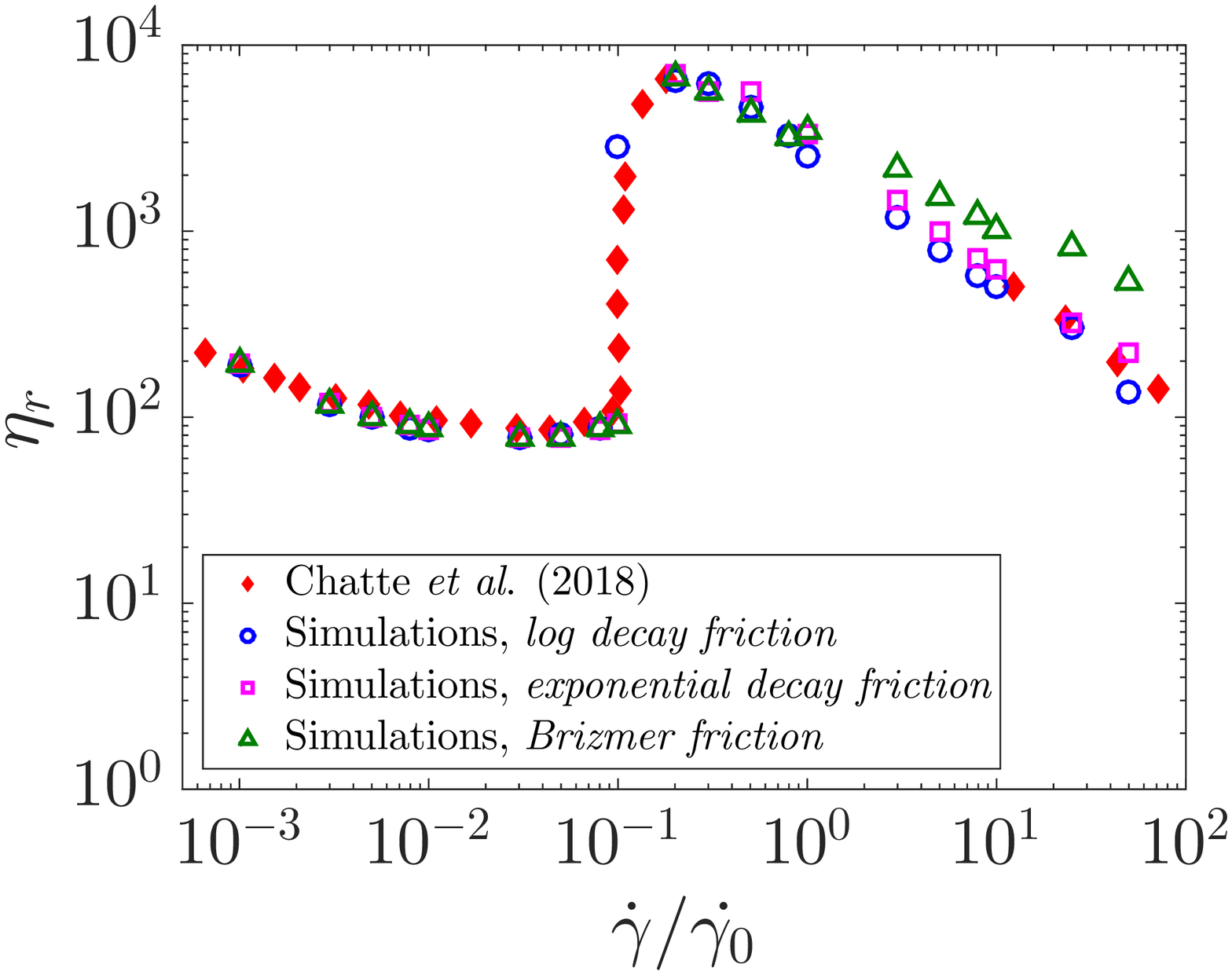}
        \caption{\label{fig:fig4b}}
    \end{subfigure}
    \caption{\label{fig:fig4}a) Contributions from hydrodynamic ($\eta^{Hydrodynamic}$), non-contact (DLVO and non-DLVO, $\eta^{Non-Contact}$) and contact ($\eta^{Contact}$) interaction to the total relative viscosity of the suspension for $\phi/\phi_{RCP} \approx 0.86$. The trends in the respective contribution follow from fig.~\ref{fig:fig3}. Lines are for guiding the eye. b) Flow curve for different coefficient of friction functions for $\phi/\phi_{RCP} \approx 0.86$.}
    \vspace{-5mm}
\end{figure}

The direct consequence of the shift in the PDF of $\langle \lambda \rangle$ can be seen on the different contributions from hydrodynamic ($\eta_r^{Hydrodynamic}$), non-contact ($\eta_r^{Non_Contact}$, DLVO and non-DLVO) and contact ($\eta_r^{Contact}$) interaction to the total relative viscosity ($\eta_r^{Total}$) in fig.~\ref{fig:fig4a}. As we increase the shear rate, $\eta_r^{Hydrodynamic}$ increases gradually. At low and intermediate shear rate values, $\eta_r^{Contact}$ is 0 as the repulsive barrier prevents direct contacts. In this regime, $\eta_r^{Non-Contact}$ decreases with increasing the shear rate which explains the first shear thinning behavior. But beyond $\dot{\gamma_c}$ the particles come into direct contacts thus resulting in the sudden jump in $\eta_r^{Total}$ due to high $\eta_r^{Contact}$. This is also known as lubricated-frictional transition which has been well studied \cite{Morris2018}. In the high shear rate regime beyond $\dot{\gamma_c}$, the contribution from the contact interactions to the bulk suspension stress is dominant and hence determines the suspension viscosity. Since $\mu$ decreases with increasing the shear rate due to lowering of $\lambda$, $\eta_r^{Contact}$ and as a consequence $\eta_r^{Total}$ decreases with an increase in the shear rate.

We also report the sensitivity of the model to different friction models in fig.~\ref{fig:fig4b}. This plot shows that the quantitative behavior of $\eta_r$ is sensitive to the exact expression for $\mu$ in the boundary lubrication regime on the Stribeck curve. But the qualitative behavior is the same for different friction models considered here, i.e., shear thinning at high shear rates. This is due to the fact that all of these friction models have a common characteristic. The $\mu$ value is high at low $|\delta|$ values when the contacts are elastic and $\mu$ decreases as the contacts become plastic. This is a fascinating result and indicates the universality of the model. The exact variation of $\mu$ in the boundary lubrication regime is particle material dependent and might have a different expression. But as long as it decreases with decrease in $\lambda$ (or an increase in $|\bm{F}_n|$) we expect to get this second shear thinning regime after the shear thickening transition at high shear rate values for dense non-Brownian suspensions. In addition, the slope of $\eta_r$ versus $\dot{\gamma}$ in the first shear thinning regime might change with changing the particle material or the suspending fluid. This can be easily incorporated in the proposed model by changing $\kappa$ and $F_A$ which are essentially material properties \cite{singh2019yielding, israelachvili2011intermolecular}.



{\textit{Conclusions.}}  We propose a unifying mechanism based on the Stribeck curve for the friction coefficient and the presence of short range repulsive force of non-DLVO origins to explain the various  rheological behaviors observed in a dense non-Brownian suspension for increasing shear rate (shear thinning - Newtonian plateau - shear thickening - shear thinning). We verify the validity of the model against experiments. The results show that the geometric constraints imposed by the range of inter-particle interactions determine the dominant interaction depending on the stress in the suspension and subsequently the transition from one regime to the other in the rheological flow curve. 
Thus, to gain further insights into the physics behind the rheological behavior of dense suspensions, accurate measurements of inter-particle forces and the coefficient of friction as a function of inter-particle gap while immersed in the fluid medium are needed. 


\begin{acknowledgments}
This research was made possible by the grants from the National Science Foundation (CBET-1604423, CBET-1705371, CBET-1700961), Department of Energy (Contract DE-EE0008256) and from Pharos Materials, Inc. \\

\end{acknowledgments}



\bibliographystyle{apsrev}

\renewcommand{\bibsection}{}

\end{document}